\documentclass{article}
\usepackage{spconf,amsmath,graphicx}
\usepackage{tabularx}
\usepackage{xcolor}

\title{FULLY LEARNABLE FRONT-END FOR MULTI-CHANNEL ACOUSTIC MODELING \\ USING SEMI-SUPERVISED LEARNING}
\name{Sanna Wager$^{1 *}$, Aparna Khare$^{2}$, Minhua Wu$^{2}$} 
\secondlinename{Kenichi Kumatani$^{2}$, Shiva Sundaram$^{2}$\thanks{*Sanna Wager performed this work as a research intern at Amazon.}}

\address{$^1$ Indiana University, School of Informatics, Computing, and Engineering, Bloomington, IN, USA\\
$^2$Amazon, Inc, Sunnyvale, CA\\}

\begin{document}
%\ninept
%
\maketitle
\begin{abstract}
In this work, we investigated the teacher-student training paradigm to train a fully learnable multi-channel acoustic model for far-field automatic speech recognition (ASR). Using a large offline teacher model trained on beamformed audio, we trained a simpler multi-channel student acoustic model used in the speech recognition system. For the student, both multi-channel feature extraction layers and the higher classification layers were jointly trained using the logits from the teacher model.  In our experiments, compared to a baseline model trained on about 600 hours of transcribed data, a relative word-error rate (WER) reduction of about 27.3\% was achieved when using an additional 1800 hours of untranscribed data. We also investigated the benefit of pre-training the  multi-channel front end to output the beamformed log-mel filter bank energies (LFBE) using L2 loss. We find that pre-training improves the  word error rate by 10.7\% when compared to a multi-channel model directly initialized with a beamformer and mel-filter bank coefficients for the front end. Finally, combining pre-training and teacher-student training produces a WER reduction of 31\% compared to our baseline. 
\end{abstract}

\section{Introduction}
\label{sec:intro}
Multi-channel far-field automatic speech recognition systems typically perform multiple tasks, including voice activity detection, speaker localization, speech enhancement, beamforming, and acoustic modeling. Beamforming is an important component at the front-end that uses spatial information from a microphone array to improve robustness against noise or reverberation, which can improve ASR accuracy \cite{wolfel2009distant}. In traditional approaches \cite{delcroix2014linear, benesty2008microphone, brandstein2013microphone, kumatani2012microphone, himawan2010clustered, sullivan1993multi}, beamforming is performed as a pre-processing step, and is not data-driven: its output is used as input to the trainable part of the acoustic model (AM). While this approach is effective, it can often fail when the room and speaker conditions do not match the design criteria. Prior work has shown that learning a multi-channel front-end jointly with the AM using the ASR objective can improve far-field performances. %\cite{sainath2017multichannel, ochiai2017multichannel, higuchi2016optimization, minhua2019frequency, kumatani2019multi}. 
In \cite{sainath2017multichannel}, Sainath et al. showed that input from a data-driven multi-channel front-end provides better results than both single-channel and beamformed input. They introduce a set of convolutional filters applied directly to the raw audio \cite{sainath2017multichannel}. The convolutional and linear structures are both designed to explicitly incorporate multiple beamformer ``look directions'', subsuming a multi-geometry beamforming component into the deep neural network (DNN). Ochiai et al. present a bi-directional LSTM structure that learns either a filter or mask estimation beamformer given frequency-domain input signals \cite{ochiai2017multichannel}. Wu et al. in \cite{minhua2019frequency} use a set of linear transformations applied to the frequency-domain input signals that also subsumed the notion of look directions. That work primarily introduced initializing the spatial filtering layers with superdirective beamformer coefficients and training the model in a stagewise manner. This was also extended to the multi-geometry case by Kumatani et al. in \cite{kumatani2019multi}.  The multi-channel spatial filtering approach described in these works allows the front-end to be trained on challenging real-world examples directly on the ASR task.

Teacher-student training (T/S) or knowledge distillation was described in  \cite{li2014learning} and \cite{hinton2015distilling}. The authors  demonstrated that the posterior probabilities generated by powerful offline ``teacher'' models can be used to train simpler ``student'' models and the technique was successfully applied to the acoustic modeling problem in \cite{li2014learning}. This technique has also been successfully applied for domain adaption for ASR. In Li et al. \cite{li2017large}, the authors  improve speech recognition performance of a distant microphone by applying T/S training to utterances recorded simultaneously using a close-talking  distant microphones.  In a similar vein, Mosner et al. \cite{movsner2019improving} apply T/S to improve noise robustness by creating a parallel corpus by adding multimedia interference to clean utterances. T/S strategy has also been used for  improving the overall  ASR performance of the student model by leveraging significantly larger amount of untranscribed  or unlabelled speech data. Parthasarathi et al. \cite{parthasarathi2019lessons} present results on using up to 1 million hours of untranscribed data. 

%\subsection{contribution}
In contrast to the prior work listed above, the main contribution of this work is the use of teacher-student training to jointly train the spatial filtering, feature extraction and classification layers. It combines and builds upon the prior work where we train a unified multi-channel acoustic model by  leveraging significantly larger amount of untranscribed data for acoustic model training. In contrast to other works that used teacher-student training for domain adaptation or knowledge distillation, we specifically focus on learning a front-end and improve the performance of a multi-channel ASR by training the model on real-world examples. We also experimentally evaluate the benefit of pre-initializing the front-end layers. Specifically, we compare initializing the spatial filtering weights with traditional signal processing based beamformer coefficients against a data-driven approach that uses L2-loss.

This paper is structured as follows. Section \ref{sec:model} introduces the network architecture used in our experiments along with initialization, pre-training approaches, and the T/S training. Section \ref{sec:results} describes the datasets and training techniques we use for our experiments, our experimental setup and the results. Finally, Section \ref{sec:conclusion} concludes the paper.

\section{Model structure}
\label{sec:model}
\subsection{Multi-channel acoustic model}
\begin{table*}
\caption{List of experiments, including the initialization used for the beamforming and MFB layers and the linear layer in-between, either random or based on a DSP technique. Each experiment is trained first using cross-entropy (Xent), then sMBR.}
\label{table:experiments}
\centering
\begin{tabularx}{7in}{ |c|X|X|c| } 
\hline
\multicolumn{4}{|c|}{\textbf{Experiment settings}}\\
\hline\hline
\textbf{Model names} & \textbf{Training dataset} & \textbf{Initialization technique} & \textbf{training technique} \\
\hline\hline
Random-init-Xent + sMBR & 621 hours & Xavier  & Supervised \\
\hline
DSP-init-Xent + sMBR & 621 hours & DSP, Xavier & Supervised \\
\hline
Pretrained-Xent + sMBR & pretrain: 1818, train: 621 & Xavier + pre-training & Supervised \\ 
\hline
Pretrained-DSP-init-Xent + sMBR &  pretrain: 1818, train: 621 & DSP, Xavier + pre-training & Supervised \\ 
\hline
Student-DSP-init-Xent + sMBR & 1818 hours & ``DSP-init-sMBR'' weights & Teacher-Student \\ 
\hline
Student-pretrained-Xent + sMBR & 1818 hours & ``Pretrained-sMBR'' weights & Teacher-Student \\ 
\hline
Teacher sMBR & 71500 hours & Uniform & Supervised \\
\hline
\end{tabularx}
\end{table*}

The model architecture used in our experiments is the \textit{elastic spatial filtering} used in \cite{minhua2019frequency}. We compute a 128-dimensional discrete Fourier transform (DFT) over a 12.5-ms window with a 10-ms frame shift, removing the DC component. After global mean-variance normalization (GMV norm.), the channels are input to the front-end component of the network, which is the combination of a beamformer and a LFBE feature extractor. 
%as described in Section \ref{sec:pretraining
Next is a complex power operation reducing the dimension to 1536, followed by a linear layer that produces a weighted combination of the spatial filtering layer outputs and reduces the dimension to 127. The feature extraction component consists of a Mel filter Bank (MFB) layer with output dimension 64, followed by the ReLU and log operations, which mimic the LFBE. Note that the various linear layers are named after their expected digital signal processing (DSP) functionality but, with training, might learn different transformations. 

\begin{figure}[t]
    \centering
    \includegraphics[width=0.4\textwidth]{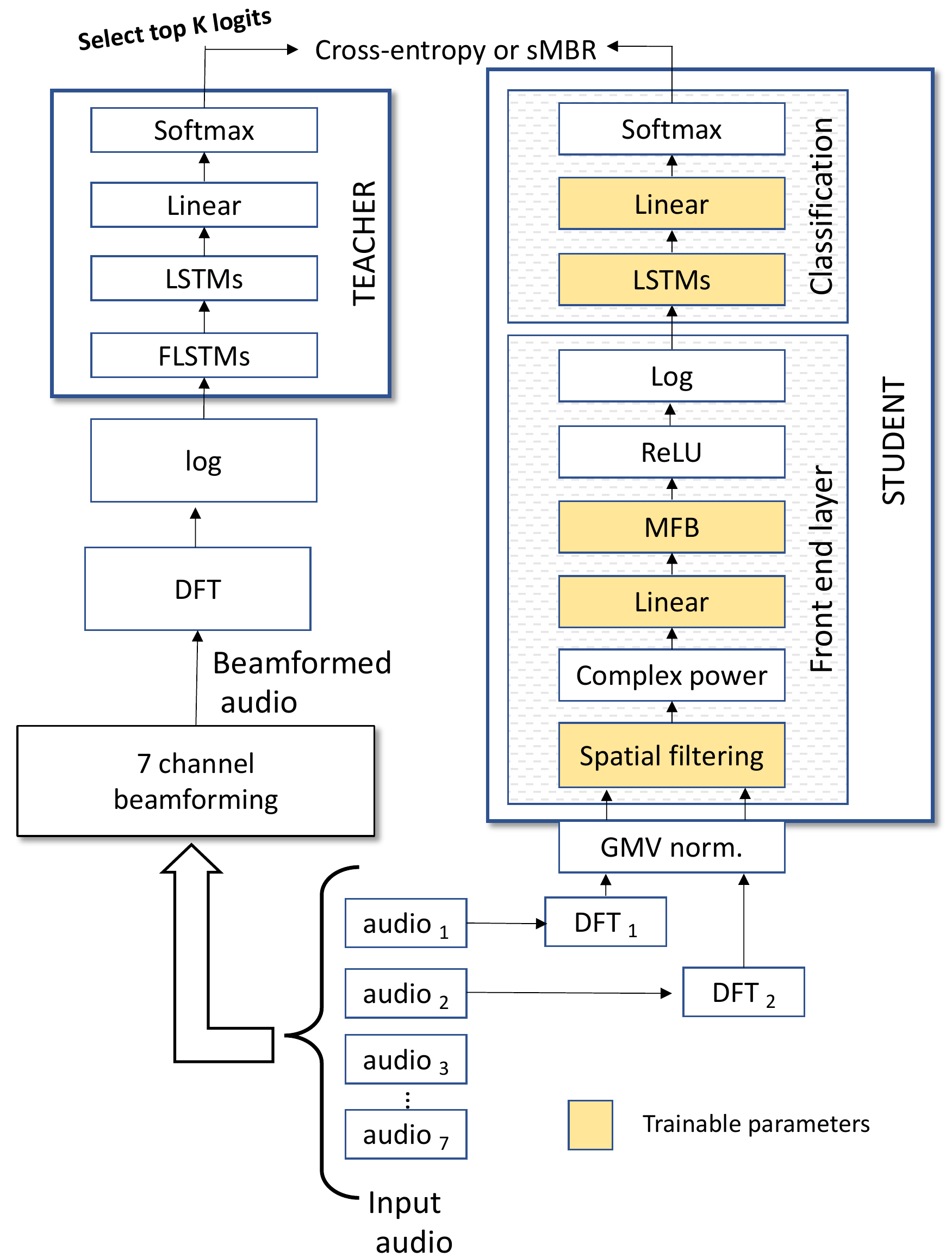}
    \caption{Teacher-student training of the multi-channel acoustic model architecture. The left side shows the teacher architecture, and the right side shows the architecture of the model used both as a standalone and as a student model.}
    \label{fig:ts_model}
\end{figure}

 The classification component of the AM contains 5 unidirectional LSTM layers with a hidden dimension of 768, followed by a linear layer and the softmax operation that outputs a 3183-dimensional senone probability distribution. The total number of weights in the network is 29.6M. The model structure is shown on the right side of Figure \ref{fig:ts_model} that describes  the overall T/S training.  The model architecture is consistent across all experiments described in this paper.

\begin{figure}[t]
    \centering
    \includegraphics[width=0.4\textwidth]{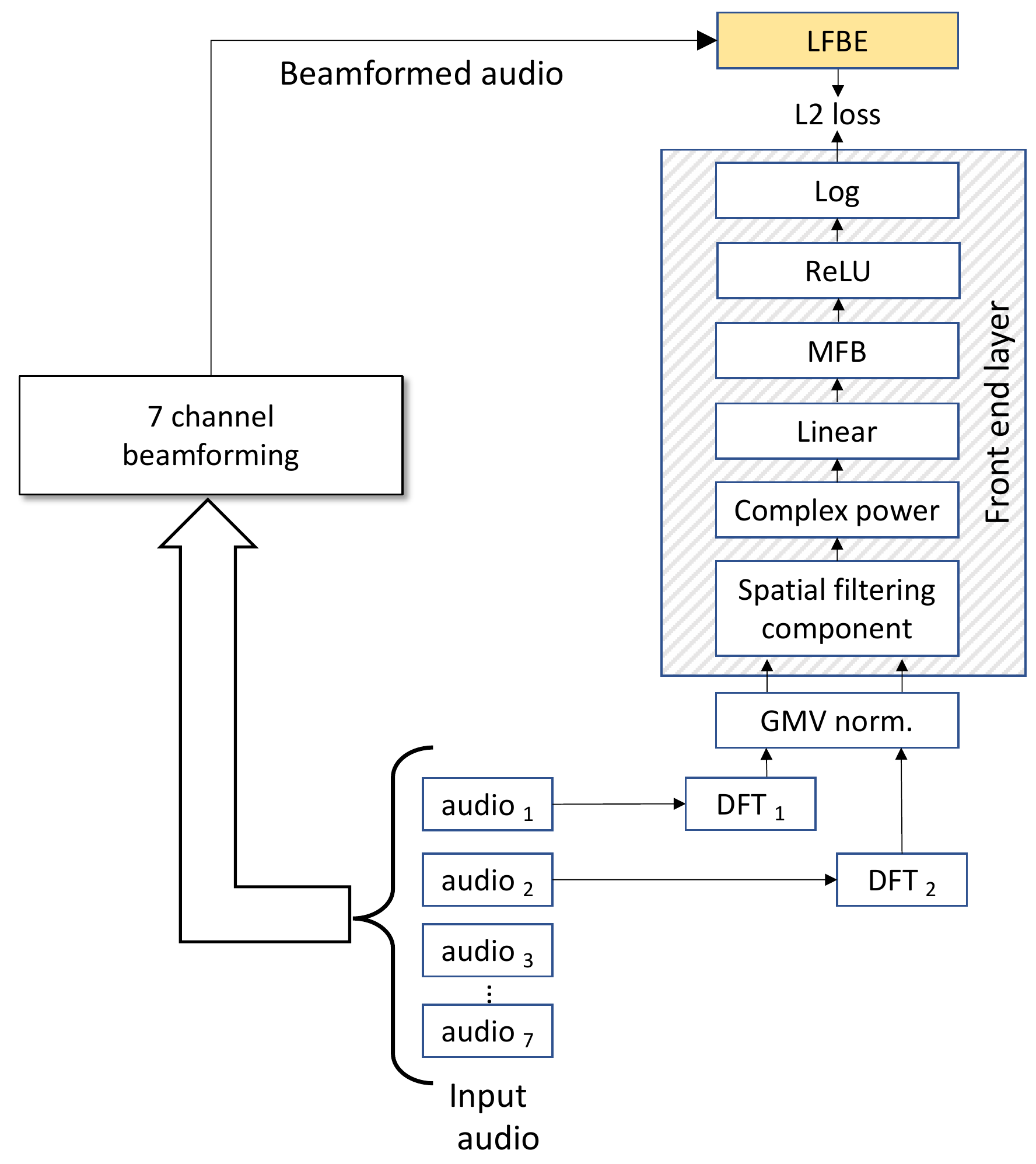}
    \caption{Pre-training of the front-end layers up to the classification network.  LFBE targets are computed using beamformed audio using 7 channels, while the multi-channel model uses only 2 out of the 7 channels}
    \label{fig:pretraining}
\end{figure}

\subsection{Initialization and pre-training}
\label{sec:pretraining}
In this work we initialize the LSTM layers with parameters trained on 1 million hours of beamformed LFBE input \cite{parthasarathi2019lessons}. As mentioned before, the main focus of this work is learning the front-end layers of the multi-channel acoustic model and the concomitant initialization.  To this end, we examine various initialization strategies for the front-end layers.

In our baseline model, the spatial filtering component is initialized with a set of 12 super-directive beamformer coefficients that use the spherically isotropic noise field \cite{doclo2007superdirective, himawan2010clustered}, each with a different look direction, as described in \cite{minhua2019frequency}. This is the `DSP-init' baseline. The output of the spatial filtering layer is then combined using an affine transform that is initialized using Xavier-normal \cite{glorot2010understanding}, and the MFB affine transform with MFB weights. For reference, we also try initializing all three layers with Xavier-normal (cf.`Random-init').

We compare the standard initializations (DSP-init and Random-init)  to data-driven pre-training of the front-end and joint training of the front-end and classification layers via T/S training. Our front end pre-training method is inspired by a minimum mean squared error beamformer approach \cite{eldar2007competitive}.  We leverage the fact that we can compute the beamformed LFBE directly from 7 channel audio by using the super-directive beamformer \footnote{For all the experiments, the beamformed audio and multi-channel audio were delay compensated.}. With the LFBE features computed from beamformed audio, we pre-train the front-end layer with 2-channel raw input using L2-loss  against the beamformed LFBE. This is shown in Figure \ref{fig:pretraining}. For pre-training  the front-end with L2-loss, we experimented with initializing the model using DSP-based and random. When using DSP-based parameters, we initialize the linear layer in-between using a random uniform distribution whose minimum and maximum are the average of the beamformer and MFB minimum and maximum values. During pre-training, we fix the DSP-based beamforming and MFB weights for the first epoch, then fine-tune them along with the other parameters. Both setups converged best with batch size 16, learning rate 0.0001, and an Adam optimizer with $\beta=(0.9, 0.999)$ and $eps=1e-08$. Interestingly, we found that the randomly initialized model converged faster and to a lower L2-loss than the model initialized with DSP-based parameters.

\begin{table*}
\centering
\begin{tabular}{|c|c|c|}
\hline
\textbf{Index} & \textbf{Model name} &  $\textbf{WER Reduction \%}$ \\
\hline
1 & \textbf{DSP-init-Xent + sMBR training} (baseline)  &  - \\
2 & + T/S training & 27.3 \\
\hline
3 & Random-init-Xent + sMBR training &  -65.3  \\
\hline
4 & Pretrained-DSP-init-Xent + sMBR training &  -5.0  \\
\hline
5 & Pretrained-Xent + sMBR training  &  10.7 \\
6 & + T/S training & 31.0 \\
\hline
7 & \textbf{Teacher-biLSTM} (offline) & 36.0  \\
\hline
\end{tabular}
\caption{Results by different initialization schemes and T/S training}
\label{table:results}
\end{table*}

\subsection{Teacher-student training}
Our teacher model trained in a supervised manner on beamformed data. T/S training is semi-supervised, as it leverages untranscribed utterances fed to the teacher in the beamformed format and to the student in MC format. Additionally, the teacher model is larger and more complex than the student whose structure is designed for low latency, which makes this instance of T/S training knowledge distillation. Soft labels should be more informative than hard labels, providing the student the ability to learn more complex functions \cite{hinton2015distilling}. We use the techniques described in prior work on T/S training described in Section \ref{sec:intro} \cite{li2017large, movsner2019improving, hinton2015distilling}. Instead of using all 3183 senone probabilities, we only use the top 20 probabilities as described in \cite{movsner2019improving}. This improves learning and saves space. We also soften the senone logits output by the teacher using temperature $T$. For all our experiments, we use $T=2$ since that was found to be the optimal parameter in \cite{movsner2019improving}.

Our teacher model is trained using beamformed 256 dimensional log-DFT features. It is then used to output logits used for teacher-student training. The teacher model architecture consists of a 2-layer bi-directional F-LSTM \cite{li2015lstm} with hidden dimension 16, frequency dimension window 48, and stride dimension 15, followed by a 5-layer bidirectional LSTM with hidden dimension 768. The total number of weights in the model is 75.9M. 
\section{Experiment setup and Results}
\label{sec:results}
All of the data for these experiments was collected using a 7-channel circular microphone array described in \cite{kumatani2019multi}. The beamformed data, used for training the teacher model and generating the LFBE targets for the L2-loss based pre-initialization, was generated using the superdirective beamformer  using the 7 channels as described in \cite{minhua2019frequency}. For all the multi-channel experiments, we selected 2 microphone channels from the opposite sides of the microphone array.
We used 621 hours of transcribed multi-channel data for supervised training. For T/S training and pre-training, we used 1818 hours of pooled transcribed and untranscribed data, which included the 621 hours. Even when the data was transcribed, we used the soft targets, not the transcriptions. The teacher model was trained on 71500 hours of transcribed beamformed data. The test set contains real-world far-field data, with a  total of 58183 utterances for analysis. The MC AM was first trained in a supervised manner using cross entropy (Xent) and State-level Minimum Bayes Risk (sMBR) training, then used to initialize the student. For sMBR training of the student model, we just used the 621 hours of supervised data.

\subsection{Results with different initialization techniques}
A summary of our experiments is displayed in Table \ref{table:experiments}. For supervised training, we compare four different initialization techniques described in Section \ref{sec:model}. The first two are initialized directly, either with DSP-based components (DSP-init, 1 in Table \ref{table:results}) or randomly (Random-init, 3 in Table \ref{table:results}). The next two add pre-training to the first two (Pretrained-DSP-init and Pretrained, 4 and 5 respectively in Table \ref{table:results}). The results are displayed in Table \ref{table:results}. We use the ``DSP-init'' model (1 in Table \ref{table:results}) as our baseline because it produces better results than random initialization of the front-end.  Pre-training on randomly initialized linear transformations showed a relative WER improvement of 10.7\%  compared to the baseline.  This result demonstrates that the front-end can be learned in a fully data-driven manner. 

We observe that the pre-training works better when initialized with random weights rather than with DSP weights, we hypothesize that this is because the conventional beamformer weights are sub-optimal for the real-data conditions and initializing the parameters with those weights prevents the model from getting to the optimal parameter set with limited data. 

\subsection{Results with T/S training}
We select the two best performing models (DSP-init and Pretrained) to initialize the student (Student-DSP-init and Student-pretrained) for T/S training and the results are showed in Table \ref{table:results}. T/S training improved the WER by 27.3\% relative compared to the ``DSP-init''  baseline and helped the model perform nearly as well as the teacher despite being smaller (2 in Table \ref{table:results}). Combining pre-training and T/S training improves the the WER by 31.0\% relative (6 in Table \ref{table:results}).  For all our experiments with initialization, we saw 13-15\% relative WER improvements with sMBR training on top of cross entropy training.  However with the T/S training the improvements were 3-5\% relative, likely because it only uses transcribed data. 

The gains we observe using T/S training are in the same range as the results reported in \cite{movsner2019improving}, and our experiments demonstrate that we can distill information from a single channel teacher model to a multi-channel student model and learn the front-end components in a data-driven manner. The second observation we make is the WER improvements with pre-training improve performance even after T/S training. This result shows us that we do not need to rely on prior knowledge in order to train these models, except for the target LFBE values from beamformed audio required to pre-train the front end weights. 

\section{Conclusions}
\label{sec:conclusion}

In this paper, we explore using semi-supervised learning for optimizing the fully learnable front-end of a multi-channel acoustic model  The teacher-student approach adopted in our experiments leverages untranscribed data and distills knowledge from a complex ``teacher'' model that uses beamformed data to a low-latency multi-channel ``student'' model. We also introduce pre-training of the front-end, learning the beamforming and feature extraction layers with the beamformed LFBEs as the target. The target is computed directly from audio, making it once again possible to harness untranscribed data. We find that both techniques improve the performance of our model. 

In this work, we have only studied these techniques applied to two channels input data. In the future, we would like to apply the same techniques to a large number of multi-channel inputs. We would also like to explore knowledge distillation for state-minimum Bayes risk training \cite{kanda2017investigation} to see if we can get more gains from the sequence training stage. Finally we would also like to explore other architectures for multi-channel acoustic modeling.

%\vfill
%\pagebreak

\vfill\pagebreak

\bibliographystyle{IEEEbib}
\bibliography{refs}

\begin{thebibliography}{10}

\bibitem{wolfel2009distant}
M.~W{\"o}lfel and J.~W. McDonough,
\newblock {\em Distant speech recognition},
\newblock Wiley Online Library, 2009.

\bibitem{delcroix2014linear}
M.~Delcroix, T.~Yoshioka, A.~Ogawa, Y.~Kubo, M.~Fujimoto, N.~Ito, K.~Kinoshita,
  M.~Espi, T.~Hori, T.~Nakatani, et~al.,
\newblock ``Linear prediction-based dereverberation with advanced speech
  enhancement and recognition technologies for the {REVERB} challenge,''
\newblock in {\em Reverb workshop}, 2014.

\bibitem{benesty2008microphone}
J.~Benesty, J.~Chen, and Y.~Huang,
\newblock {\em Microphone array signal processing}, vol.~1,
\newblock Springer Science \& Business Media, 2008.

\bibitem{brandstein2013microphone}
M.~Brandstein and D.~Ward,
\newblock {\em Microphone arrays: signal processing techniques and
  applications},
\newblock Springer Science \& Business Media, 2013.

\bibitem{kumatani2012microphone}
K.~Kumatani, T.~Arakawa, K.~Yamamoto, J.~Mc{D}onough, B.~Raj, R.~Singh, and
  I.~Tashev,
\newblock ``Microphone array processing for distant speech recognition: Towards
  real-world deployment,''
\newblock in {\em Proc. Asia Pacific Signal and Information Processing
  Association Annual Summit and Conference (APSIPA ASC)}. IEEE, 2012, pp.
  1--10.

\bibitem{himawan2010clustered}
Ivan Himawan, Iain McCowan, and Sridha Sridharan,
\newblock ``Clustered blind beamforming from ad-hoc microphone arrays,''
\newblock {\em Transactions Audio, Speech, and Language Processing (TASLP)},
  vol. 19, no. 4, pp. 661--676, 2010.

\bibitem{sullivan1993multi}
T.~M. Sullivan and R.~M. Stern,
\newblock ``Multi-microphone correlation-based processing for robust speech
  recognition,''
\newblock in {\em Int. Conf. Acoustics, Speech, and Signal Processing
  (ICASSP)}. IEEE, 1993, vol.~2, pp. 91--94.

\bibitem{sainath2017multichannel}
T.~N. Sainath, R.~J. Weiss, K.~W. Wilson, B.~Li, A.~Narayanan, E.~Variani,
  M.~Bacchiani, I.~Shafran, A.~Senior, K.~Chin, et~al.,
\newblock ``Multichannel signal processing with deep neural networks for
  automatic speech recognition,''
\newblock {\em Transactions Audio, Speech, and Language Processing (TASLP)},
  vol. 25, no. 5, pp. 965--979, 2017.

\bibitem{ochiai2017multichannel}
T.~Ochiai, S.~Watanabe, T.~Hori, and J.~R. Hershey,
\newblock ``Multichannel end-to-end speech recognition,''
\newblock in {\em Proc. 34th Int. Conf. Machine Learning (ICML)}. JMLR, 2017,
  vol.~70, pp. 2632--2641.

\bibitem{minhua2019frequency}
M.~Wu, K.~Kumatani, S.~Sundaram, N.~Str{\"o}m, and B.~Hoffmeister,
\newblock ``Frequency domain multi-channel acoustic modeling for distant speech
  recognition,''
\newblock in {\em Int. Conf. Acoustics, Speech and Signal Processing (ICASSP)}.
  IEEE, 2019, pp. 6640--6644.

\bibitem{kumatani2019multi}
K.~Kumatani, M.~Wu, S.~Sundaram, N.~Str{\"o}m, and B.~Hoffmeister,
\newblock ``Multi-geometry spatial acoustic modeling for distant speech
  recognition,''
\newblock in {\em Int. Conf. Acoustics, Speech and Signal Processing (ICASSP)}.
  IEEE, 2019, pp. 6635--6639.

\bibitem{li2014learning}
Jinyu Li, Rui Zhao, Jui-Ting Huang, and Yifan Gong,
\newblock ``Learning small-size dnn with output-distribution-based criteria,''
\newblock in {\em Fifteenth annual conference of the international speech
  communication association}, 2014.

\bibitem{hinton2015distilling}
G.~Hinton, O.~Vinyals, and J.~Dean,
\newblock ``Distilling the knowledge in a neural network,''
\newblock {\em stat}, vol. 1050, pp. 9, 2015.

\bibitem{li2017large}
J.~Li, M.~L. Seltzer, X.~Wang, R.~Zhao, and Y.~Gong,
\newblock ``Large-scale domain adaptation via teacher-student learning,''
\newblock {\em Proc. Interspeech}, 2017.

\bibitem{movsner2019improving}
L.~Mo{\v{s}}ner, M.~Wu, A.~Raju, S.~H.~K. Parthasarathi, K.~Kumatani,
  S.~Sundaram, R.~Maas, and B.~Hoffmeister,
\newblock ``Improving noise robustness of automatic speech recognition via
  parallel data and teacher-student learning,''
\newblock in {\em Int. Conf. Acoustics, Speech and Signal Processing (ICASSP)}.
  IEEE, 2019, pp. 6475--6479.

\bibitem{parthasarathi2019lessons}
S.~H.~K. Parthasarathi and N.~Strom,
\newblock ``Lessons from building acoustic models with a million hours of
  speech,''
\newblock in {\em Int. Conf. Acoustics, Speech and Signal Processing (ICASSP)}.
  IEEE, 2019, pp. 6670--6674.

\bibitem{doclo2007superdirective}
S.~Doclo and M.~Moonen,
\newblock ``Superdirective beamforming robust against microphone mismatch,''
\newblock {\em Transactions Audio, Speech, and Language Processing (TASLP)},
  vol. 15, no. 2, pp. 617--631, 2007.

\bibitem{glorot2010understanding}
X.~Glorot and Y.~Bengio,
\newblock ``Understanding the difficulty of training deep feedforward neural
  networks,''
\newblock in {\em Proc. 13th Int. Conf. Artificial Intelligence and Statistics
  (AISTATS)}, 2010, pp. 249--256.

\bibitem{eldar2007competitive}
Yonina~C Eldar, Arye Nehorai, and Patricio~S La~Rosa,
\newblock ``A competitive mean-squared error approach to beamforming,''
\newblock {\em IEEE Transactions on Signal Processing}, vol. 55, no. 11, pp.
  5143--5154, 2007.

\bibitem{li2015lstm}
Ji. Li, A.~Mohamed, G.~Zweig, and Y.~Gong,
\newblock ``Lstm time and frequency recurrence for automatic speech
  recognition,''
\newblock in {\em Workshop Automatic Speech Recognition and Understanding
  (ASRU)}. IEEE, 2015, pp. 187--191.

\bibitem{kanda2017investigation}
N.~Kanda, Y.~Fujita, and K.~Nagamatsu,
\newblock ``Investigation of lattice-free maximum mutual information-based
  acoustic models with sequence-level kullback-leibler divergence,''
\newblock in {\em Automatic Speech Recognition and Understanding Workshop
  (ASRU)}. IEEE, 2017, pp. 69--76.

\end{thebibliography}

\end{document}